\documentclass[aps,prx,citeautoscript,twocolumn,longbibliography,superscriptaddress]{revtex4-1}

\usepackage[T1]{fontenc}
\usepackage{latexsym}
\usepackage{graphicx} 
\usepackage{epstopdf}
\usepackage{amsmath}  
\usepackage{amssymb}
\usepackage{comment}
\usepackage{cases}
\usepackage{lipsum}
\usepackage{float}
\usepackage{tikz}
\usepackage{makecell}
\usepackage{bbold}
\usepackage{soul}
\usepackage{color}
\usepackage[breaklinks=true]{hyperref}

\begin{document}

\title{Electron-hole asymmetry of quantum collective excitations in high-$T_c$ copper oxides}

\author{Maciej Fidrysiak}
\email{maciej.fidrysiak@uj.edu.pl}
\affiliation{Institute of Theoretical Physics, Jagiellonian University, ul. {\L}ojasiewicza 11, 30-348 Krak{\'o}w, Poland }

\begin{abstract}
  We carry out a systematic study of collective spin- and charge excitations for the canonical single-band Hubbard, $t$-$J$-$U$, and $t$-$J$ models of high-temperature copper-oxide superconductors, both on electron- and hole-doped side of the phase diagram. Recently developed variational wave function approach, combined with the expansion in inverse number of fermionic flavors, is employed. All three models exhibit a substantial electron-hole asymmetry of magnetic excitations, with a robust paramagnon emerging for hole-doping, in agreement with available resonant inelastic $x$-ray scattering data for the cuprates. The $t$-$J$ model yields additional high-energy peak in the magnetic spectrum that is not unambiguously identified in spectroscopy. For all considered Hamiltonians, the dynamical charge susceptibility contains a coherent mode for both hole- and electron doping, with overall bandwidth renormalization controlled by the on-site Coulomb repulsion. Away from the strong-coupling limit, the antiferromagnetic ordering tendency is more pronounced on electron-doped side of the phase diagram.
\end{abstract}

\maketitle
\section{Introduction}

High temperature superconductivity in layered copper oxides may be induced by either hole ({h}) or electron ({e}) doping of the parent antiferromagnetic (AF) insulating state \cite{SpalekPhysRep2022}. Nonetheless, the equilibrium phase diagrams of {e}- and {h}-doped cuprates exhibit substantial asymmetry, whose one of the most evident manifestations is broader regime of the AF-phase appearance on the e-doped side \cite{ArmitageRevModPhys2010}. Thanks to developments in spectroscopic techniques, particularly resonant inelastic $x$-ray scattering (RIXS), this static picture has been recently supplemented with a detailed account of non-equilibrium properties, including collective excitation spectra \cite{LeTaconNatPhys2011,DeanPhysRevLett2013,DeanNatMater2013,DeanPhysRevB2013,LeTaconPhysRevB2013,IshiiNatCommun2014,LeeNatPhys2014,DeanPhysRevB2014,GuariseNatCommun2014,JiaNatCommun2014,WakimotoPhysRevB2015,MinolaPhysRevLett2015,PengPhysRevB2015,EllisPhysRevB2015,HuangSciRep2016,GretarssonPhysRevLett2016,MinolaPhysRevLett2017,IvashkoPhysRevB2017,MeyersPhysRevB2017,ChaixPhysRevB2018,Robarts_arXiV_2019,FumagalliPhysRevB2019,PengPhysRevB2018,RobartsArXiV2020,IshiiPhysRevB2017,NagPhysRevLett2020,SinghPhysRevB2022,HeptingNature2018,IshiiJPhysSocJapan2019,LinNPJQuantMater2020}. The low-energy dynamics of parent AF state is dominated by spin-waves of localized spins, evolving with doping into broadened magnetic excitations of the paramagnetic state (paramagnons). Counter-intuitively, the paramagnons appear to be \emph{more robust} on the h-doped side, where AF phase is rapidly suppressed away from half-filling. RIXS has been also utilized to probe collective charge (plasmon) excitations that are observed in both hole- \cite{IshiiPhysRevB2017,NagPhysRevLett2020,SinghPhysRevB2022} and electron-doped \cite{HeptingNature2018,IshiiJPhysSocJapan2019,LinNPJQuantMater2020} cuprates. Since magnetic and charge fluctuations have been considered as candidates for the pairing mechanism in high-$T_c$ superconductors (see, e.g., \cite{ChubukovChapter2003,ScalapinoRevModPhys2012,LeTaconNatPhys2011,YuPhysRevX2020}), identification of a microscopic origin of their distinct degree of e-h asymmetry is now in demand.

Parent compounds of high-$T_c$ copper-oxides are not canonical Mott insulating systems, since they may be regarded as charge transfer insulators within the Zaanen-Sawatzky-Allen scheme \cite{ZaanenPhysRevLett1985}. In effect, the e-h asymmetry manifests itself directly at the atomic-orbital level. Nominally, the doped electrons position on the copper $3d_{x^2-y^2}$ states, while holes locate predominantly on $2p_\sigma$ oxygen orbitals. Yet, there is a systematic variation of $d$-$p$-electron redistribution between different cuprate families \cite{RybickiNatCommun2016}. Whereas minimally the $d$-$p$ model of $\mathrm{CuO_2}$ plane is needed to address those orbital-selective effects \cite{ZegrodnikPhysRevB2019,ZegrodnikEurPhysJB2020,ZegrodnikJPCM2021,LiCommunPhys2021}, doping dependence of the principal experimental equilibrium properties and their e-h asymmetry is reproduced already within the effective single-band picture \cite{SpalekPhysRevB2017,ZegrodnikPhysRevB2017,FidrysiakJPhysCondensMatter2018,WysokinskiPhysRevB2017}. The latter model is also appropriate for studies of the  correlated state dynamics following interaction quench \cite{SchiroPhysRevLett2010,WysokinskiPhysRevB2017_2}. The single-band Hamiltonians remain thus a common starting point for describing the physics of high-$T_c$ cuprates and related materials.

Here we carry out a study of the e-h asymmetry of collective magnetic and charge excitations for the canonical Hubbard, $t$-$J$-$U$, and $t$-$J$ models of high-$T_c$ copper-oxide superconductors, which allows us to single out the microscopic parameters controlling relevant aspects of their many-particle dynamics. The recently introduced concept of effective exchange interaction \cite{FidrysiakJMMM2021} is invoked to relate all three Hamiltonians in proper limits so that their collective excitations can be directly compared and analyzed on equal footing. In order to account for the effects of strong electronic correlations, we supplement \textbf{V}ariational \textbf{W}ave \textbf{F}unction (VWF) approach with the expansion in inverse number of fermionic flavors ($1/\mathcal{N}_f$), which results in VWF+$1/\mathcal{N}_f$ scheme \cite{FidrysiakPhysRevB2021}. The latter has been demonstrated to be effective in semi-quantitative analysis of observed spin- and charge dynamics of h-doped cuprates, and to compare favorably with other computational techniques, including determinant quantum Monte-Carlo \cite{FidrysiakPhysRevB2020,FidrysiakPhysRevBLett2021,FidrysiakJMMM2021,SpalekPhysRep2022}.

For the cases of Hubbard and $t$-$J$-$U$ models up to moderately large value of on-site interaction, we find that intense dispersive peak in magnetic spectrum persists along the anti-nodal ($\Gamma$-$X$) Brillouin-zone direction down to heavily h-overdoped regime, whereas the paramagnons along the nodal ($\Gamma$-$M$) line are rapidly suppressed. On the other hand, spin excitations are shifted to larger energies with e-doping and become less coherent away from half-filling. These findings indicate a substantial e-h asymmetry at the dynamical level and are in agreement with available RIXS data for the cuprates. Moreover, we find that the magnetic spectra calculated in the strong-coupling limit ($t$-$J$ model) differ significantly from those obtained for the Hubbard and $t$-$J$-$U$ Hamiltonians, and exhibit characteristics that are not observed experimentally. This might suggest that the $t$-$J$ model overestimates local electronic correlations, as has been also noted within former theoretical survey of equilibrium and single-particle quantities \cite{SpalekPhysRep2022}.

For completeness, we also address the doping evolution of collective charge excitations that are demonstrated to persist on both h- and e-doped sides of the phase diagram. The dynamical charge response unambiguously separates into the incoherent continuum part and a coherent charge mode. With increasing on-site Coulomb repulsion, $U$, the collective mode energy undergoes a systematic downward renormalization. Thus, the electronic band-narrowing effects are instrumental for the charge sector, whereas magnetic part remains primarily sensitive to effective exchange interaction, $J_\mathrm{eff}$. Remarkably, within the $t$-$J$-$U$ model, the parameters $U$ and $J_\mathrm{eff}$ remain independent and may be simultaneously tuned to match experiment. This is not possible within the Hubbard and $t$-$J$ models. Such flexibility is needed to account for non-trivial superexchange pathways via oxygen in a charge-transfer insulator that alter the correspondence between kinetic exchange and $U$ \cite{SpalekPhysRevB2017}. 

Finally, we carry out analysis of the leading ordering instabilities of the paramagnetic state both against spin- and charge fluctuations. For the Hubbard and $t$-$J$-$U$ models, we find that the static magnetic susceptibility at the $M$ Brillouin-zone point is enhanced for e-doping relative to the h-doped side, implying stronger tendency towards antiferromagnetism. This reflects the experimentally observed e-h asymmetry of the equilibrium AF order.

\section{Model and method}
\label{sec:model}

We employ a generic square-lattice $t$-$J$-$U$ model of the copper-oxygen plane, given by the Hamiltonian

\begin{align}
  \hat{\mathcal{H}} = \sum_{i \neq j} t_{ij} \hat{a}^\dagger_{i\sigma} \hat{a}_{j\sigma} + J \sum_{\langle i, j \rangle} \hat{\mathbf{S}}_i \cdot \hat{\mathbf{S}}_j + U \sum_i \hat{n}_{i\uparrow} \hat{n}_{i\downarrow},
  \label{eq:tju-model}
\end{align}

\noindent
where $\hat{a}_{i\sigma}$ ($\hat{a}^\dagger_{i\sigma}$) are annihilation (creation) operators of spin-$\sigma$ electrons on site $i$, $\hat{n}_{i\sigma} \equiv \hat{a}^\dagger_{i\sigma} \hat{a}_{i\sigma}$, and $\hat{\mathbf{S}}_i \equiv (\hat{S}^x_i, \hat{S}^y_i, \hat{S}^z_i)$ denotes local spin operator. Out of the hopping integrals, $t_{ij}$, we retain only those connecting nearest- and next-nearest neighbors, $t \equiv -0.35\,\mathrm{eV}$ and $t^\prime \equiv 0.25 |t|$, respectively. It should be remarked that, within the $t$-$J$-$U$ model, the degree of e-h asymmetry is controlled exclusively by the magnitude of next-nearest neighbor hopping, $t^\prime$. Up to a trivial energy and chemical potential shift, the remaining terms remain unchanged after application of the e-h transformation in the usual form

\begin{align}
  a^\dagger_{i\sigma} \rightarrow a_{i\sigma} \cdot (-1)^i,  a_{i\sigma} \rightarrow a_{i\sigma}^\dagger \cdot (-1)^i,
  \label{eq:ph_transformation}
\end{align}

\noindent
with $(-1)^i$ being phase factor alternating between neighboring sites. The electronic interactions are governed by the magnitude of on-site Coulomb repulsion, $U$, and by nearest-neighbor AF exchange integral, $J$.

The $t$-$J$-$U$ Hamiltonian~\eqref{eq:tju-model} may be regarded as a generalization of the Hubbard- and $t$-$J$ models, and encompasses both of them as particular cases. The Hubbard model is obtained for $U > 0$ and $J = 0$, whereas the $t$-$J$ model for $U = \infty$ and $J > 0$. In order to analyze all three Hamiltonians on equal footing, we refer to the concept of \emph{effective exchange interaction} $J_\mathrm{eff} \equiv J + \frac{4 t^2}{U}$ that combines direct Heisenberg-type interaction with second-order kinetic exchange \cite{FidrysiakJMMM2021}. Hereafter its value is set to $J_\mathrm{eff} \equiv \frac{2}{3} |t|$ which has been used previously to study collective dynamics in the paramagnetic state of h-doped cuprates \cite{FidrysiakPhysRevBLett2021}. The selected representative parameter values are: $U = 6 |t|$, $J = 0$ (Hubbard model), $U = 12 |t|$, $J = \frac{1}{3} |t|$ ($t$-$J$-$U$ model), and $U = \infty$, $J = \frac{2}{3} |t|$ ($t$-$J$ model).

\begin{figure*}
  \centering
    \includegraphics[width=1\linewidth]{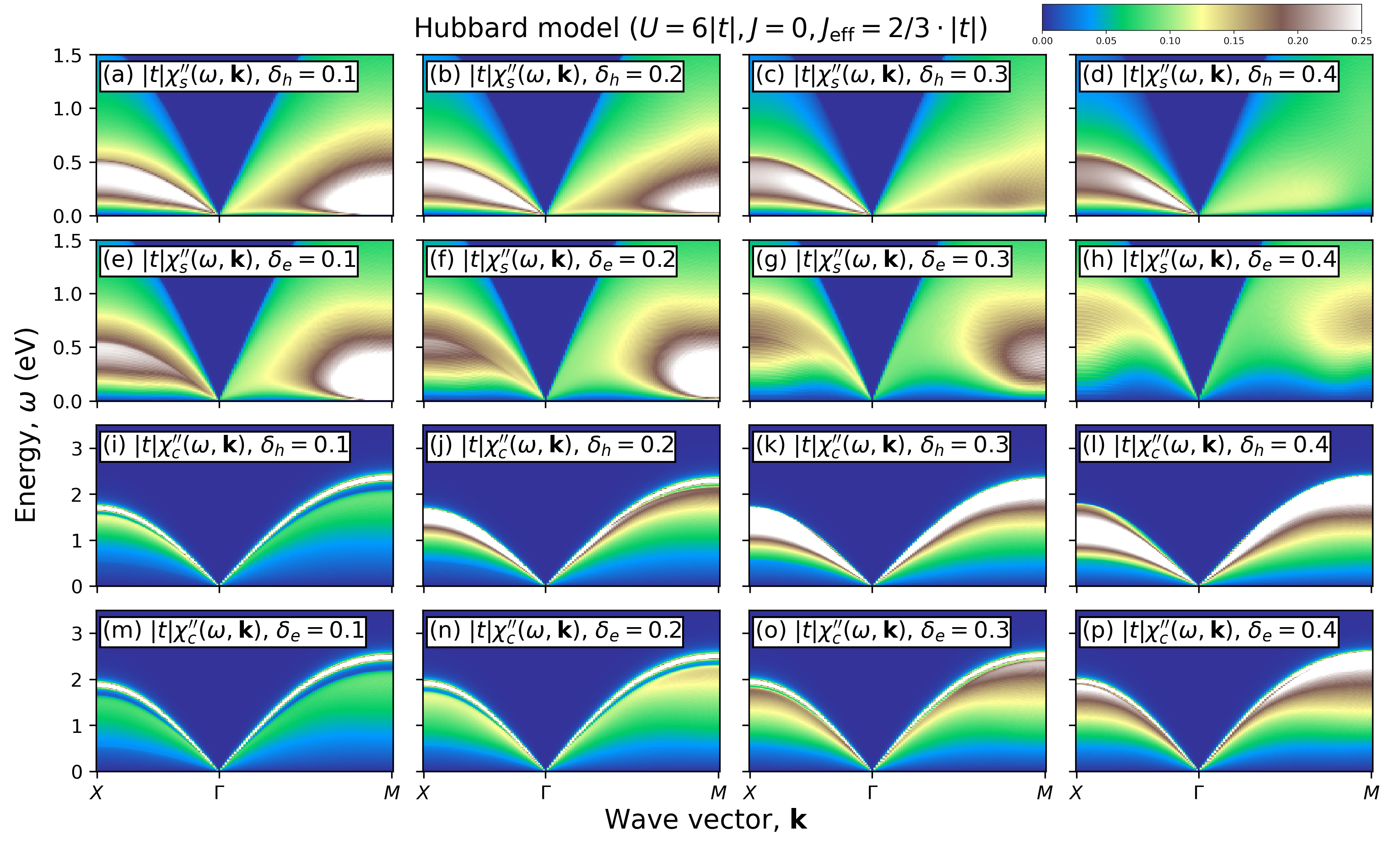}
    \caption{Calculated VWF+$1/\mathcal{N}_f$ collective excitation spectra for the Hubbard model ($t = -0.35\,\mathrm{eV}$, $t^\prime = 0.25 |t|$, $U = 6 |t|$, $J = 0$) along the high-symmetry $X$-$\Gamma$-$M$ Brillouin-zone contour. Imaginary parts of respective dynamical susceptibilities are represented as a color map, with blue- and white colors corresponding to low- and high intensity, respectively. The color scale is used consistently, so the intensities can be compared. Panels (a)-(d) and (e)-(h) show the magnetic excitations for the hole- an electron-doped system, respectively. Panels (i)-(p) illustrate the charge response, calculated for the same model parameters. The h- and e-doping levels ($\delta_h$ and $\delta_e$), detailed inside the plot, are measured relative to half-filling. }
  \label{fig:hubbard_panels}
\end{figure*}

We employ VWF+$1/\mathcal{N}_f$ approach in the local-diagrammatic variant ($\mathrm{LD}_f+1/\mathcal{N}_f$ in the notation of Ref.~\cite{FidrysiakPhysRevB2021}) that allows us to study collective excitations around the strongly-correlated ground state. The saddle-point reference point is constructed based on the variational wave function

\begin{align}
  \label{eq:psi_var}
|\Psi_\mathrm{var}\rangle \equiv \frac{\hat{P} |\Psi_0\rangle}{{|| \hat{P} |\Psi_0\rangle ||}},
\end{align}

\noindent
where $|\Psi_0\rangle$ represents a Slater determinant, and $\hat{P}$ is an operator introducing correlations into the trial state (the so-called correlator). The denominator in Eq.~\eqref{eq:psi_var} is needed for normalization, since $\hat{P}$ is not unitary in general. We take $\hat{P} \equiv \prod_i \hat{P}_i$ with

\begin{align}
  \label{eq:correlator}
  \hat{P}_i \equiv \lambda^0_{ i} |0\rangle_i {}_i\langle 0 | + \sum_{\sigma\sigma^\prime} \lambda^{\sigma\sigma^\prime}_{i} |\sigma\rangle_i {}_i\langle \sigma^\prime | + \lambda^d_{i} |d\rangle_i {}_i\langle d |,
\end{align}

\noindent
where states $|0\rangle_i$, $|{\uparrow}\rangle_i$, $|{\downarrow}\rangle_i$, $|d\rangle_i \equiv|{\uparrow\downarrow}\rangle_i$ span the local Hilbert space for the lattice site $i$, and $\sigma, \sigma^\prime = \uparrow, \downarrow$ enumerate spin configurations. At zero temperature, both the variational parameters $\boldsymbol{\lambda} = \bigcup_i \{\lambda^0_i, \lambda^{\uparrow\uparrow}_i, \lambda^{\uparrow\downarrow}_i, \lambda^{\downarrow\uparrow}_i, \lambda^{\downarrow\downarrow}_i, \lambda^{d}_i \}$ and wave function $|\Psi_0\rangle$ are determined by optimization of the energy functional $E_\mathrm{var}(|\Psi_0\rangle, \boldsymbol{\lambda}) \equiv \langle \Psi_\mathrm{var}| \hat{\mathcal{H}} |\Psi_\mathrm{var}\rangle$, subjected to the requirement of fixed total particle number and other constraints. Here we actually employ a thermal generalization of this variational scheme based on free energy functional \cite{FidrysiakPhysRevB2021}, with temperature set to $k_B T = 0.42 |t|$ in order to stay clear of density-wave orders. Stability of the paramagnetic state for all considered parameter sets is demonstrated in Appendix~\ref{appendix:stability}, where we also identify leading ordering tendencies. It should be remarked that the transformation~\eqref{eq:ph_transformation} preserves the variational space spanned by the parameters of the correlator~\eqref{eq:correlator}. The present methodology is thus unbiased and suitable for the analysis of intrinsic e-h asymmetry of collective excitations, as detailed in Appendix~\ref{appendix:correlator}.

In essence, VWF+$1/\mathcal{N}_f$ approach relies on promoting all the variational parameters to imaginary-time dynamical quantities, extending the number of fermionic families from one to $\mathcal{N}_f$, and subsequently expanding the variationally determined free energy functional in the powers of $1/\mathcal{N}_f$. We emphasize that, even in the employed here large-$\mathcal{N}_f$ approximation, the residual interactions between the Landau quasiparticles are incorporated. The latter are instrumental for a proper description of collective paramagnon and plasmon excitations in the strong-coupling regime.

Finally, we note that the strong-coupling limit ($U = \infty$, $J > 0$) is described within our formalism on the same footing as the finite-$U$ case. The exclusion of doubly occupied sites at the dynamical level is achieved by imposing the local constraint $\lambda^d_{i}(\tau) \equiv 0$, where $\tau$ denotes imaginary time. Since e-doped side of the phase diagram is inaccessible within the low-energy sector of the $t$-$J$-model, in this regime we actually carry out the analysis for the h-doped system with inverted sign of next-nearest hopping, $t^\prime$, as dictated by e-h transformation~\eqref{eq:ph_transformation}.

We calculate the (imaginary-time) dynamical susceptibilities in the form $\chi_s(\tau, \mathbf{k}) \equiv \langle \mathcal{T}_\tau \hat{{S}}_\mathbf{k}^z(\tau) \hat{{S}}_{-\mathbf{k}}^z  \rangle$ and $\chi_c(\tau, \mathbf{k}) \equiv \langle \mathcal{T}_\tau \hat{n}_\mathbf{k}(\tau) \hat{n}_{-\mathbf{k}}  \rangle$. The results have been obtained for a $300 \times 300$ square lattice with periodic boundary conditions, and analytic continuation of the Fourier transformed susceptibilities, $\chi_{s/c}(i \omega_n, \mathbf{k})$, is performed as $i\omega_n \rightarrow \omega + 0.02 |t|$, where $\omega_n$ are bosonic Matsubara frequencies.

\section{Results}
\label{sec:results}

\begin{figure*}
  \centering
    \includegraphics[width=1\linewidth]{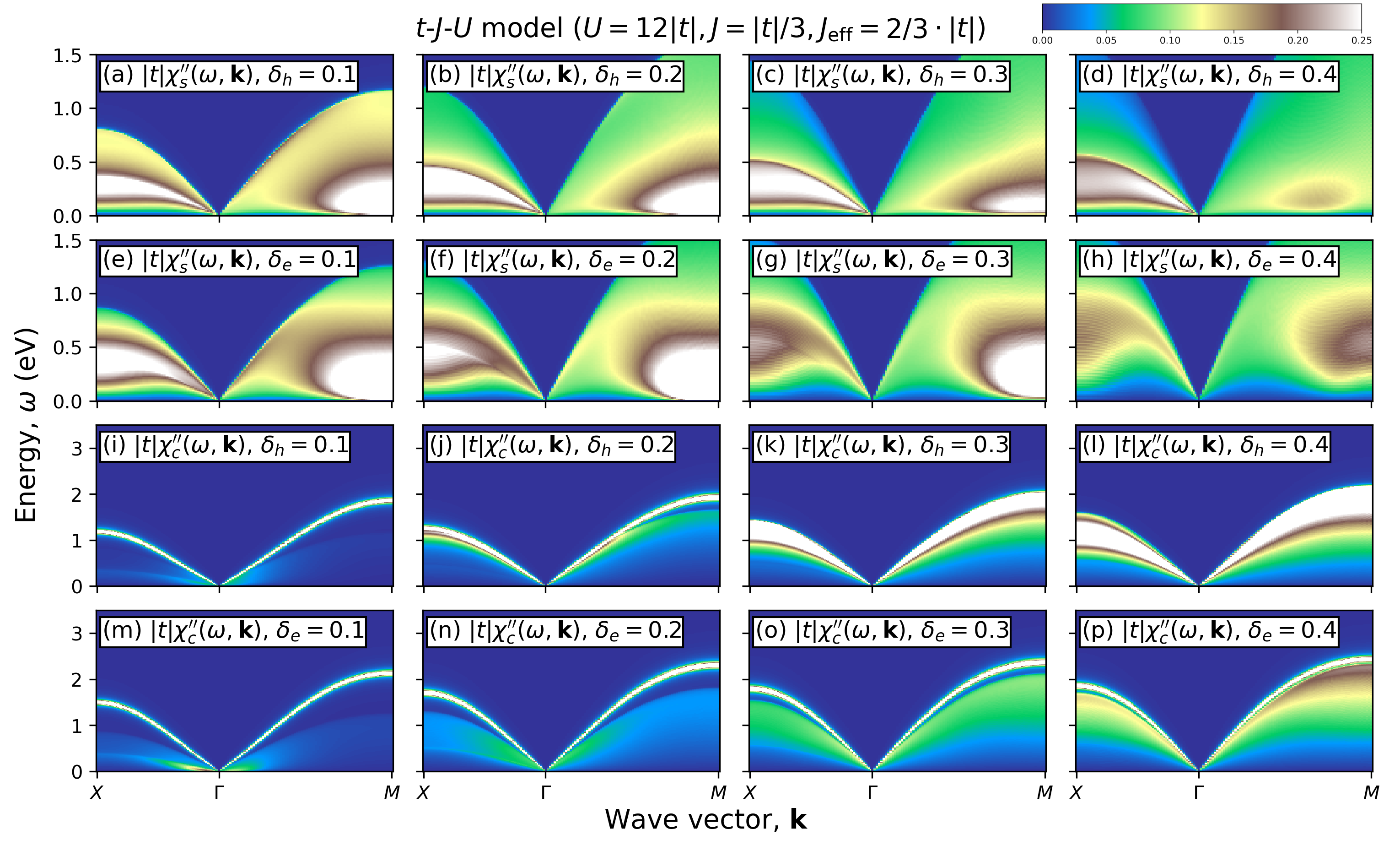}
  \caption{Summary of calculated VWF+$1/\mathcal{N}_f$ collective excitation spectra for the $t$-$J$-$U$ model ($t = -0.35\,\mathrm{eV}$, $t^\prime = 0.25 |t|$, $U = 12 |t|$, $J = \frac{1}{3} |t|$). The plot is arranged in the same way as Fig.~\ref{fig:hubbard_panels} and the color scale is consistent among panels.}
  \label{fig:tju_panels}
\end{figure*}

\subsection{Hubbard model}
\label{sec:hubbard_model}

We start the discussion of e-h asymmetry of collective excitations with the Hubbard model case ($U = 6 |t|$, $J = 0$).  In Fig.~\ref{fig:hubbard_panels}, the calculated VWF+$1/\mathcal{N}_f$ dynamical spin [panels (a)-(d)] and charge [panels (e)-(h)] susceptibilities for hole (electron) doping levels $\delta_h  (\delta_e) = 0.1, 0.2, 0.3, 0.4$ are displayed along the high symmetry $X$-$\Gamma$-$M$ Brillouin-zone contour. The color map represents susceptibility magnitude with blue- and white colors mapping to low- and high-intensities, respectively. Relevant model parameters are given inside the panels. 

For h-doping (a)-(e), an intense and disperse signal emerges throughout entire phase diagram along the anti-nodal ($\Gamma$-$X$) direction. This peak is interpreted as robust paramagnon in high-$T_c$ literature. On the other hand, the spectrum remains highly incoherent along the nodal ($\Gamma$-$M$) line. Such a directional anisotropy of spin excitations is in agreement with RIXS data collected for  hole-doped copper oxides \cite{RobartsPhysRevB2019}.

We now proceed to comparison of the h- and e-side of the phase diagram. As follows form Fig.~\ref{fig:hubbard_panels}(e)-(h), a clear resonant peak appears for e-doping only along the $\Gamma$-$X$ direction and for $\delta_e = 0.1$ [panel (e)]. Yet, it is less intense than the corresponding feature on the h-side [panel (a)]. With increasing electron concentration, the magnetic spectral weight shifts to larger energies and rapidly becomes incoherent. Such a hardening effect has been reported experimentally for e-doped cuprates \cite{LeeNatPhys2014,IshiiNatCommun2014}. The substantial e-h asymmetry of spin excitations is thus reproduced already within the single-band Hubbard model.

For completeness, in Fig.~\ref{fig:hubbard_panels}(i)-(p), we also compare the h- and e-doping evolution of charge excitations. The spectrum unambiguously separates into the incoherent continuum part and sharp disperse collective mode. The latter remains gapless at the $\Gamma$ point, since the long-range Coulomb interactions are not included in the Hamiltonian~\eqref{eq:tju-model}. Contrary to the spin part, the charge modes emerge for both e- and h-doping, which is also backed by RIXS experiments \cite{IshiiPhysRevB2017,NagPhysRevLett2020,SinghPhysRevB2022,HeptingNature2018,IshiiJPhysSocJapan2019,LinNPJQuantMater2020}.

\begin{figure*}
  \centering
    \includegraphics[width=1\linewidth]{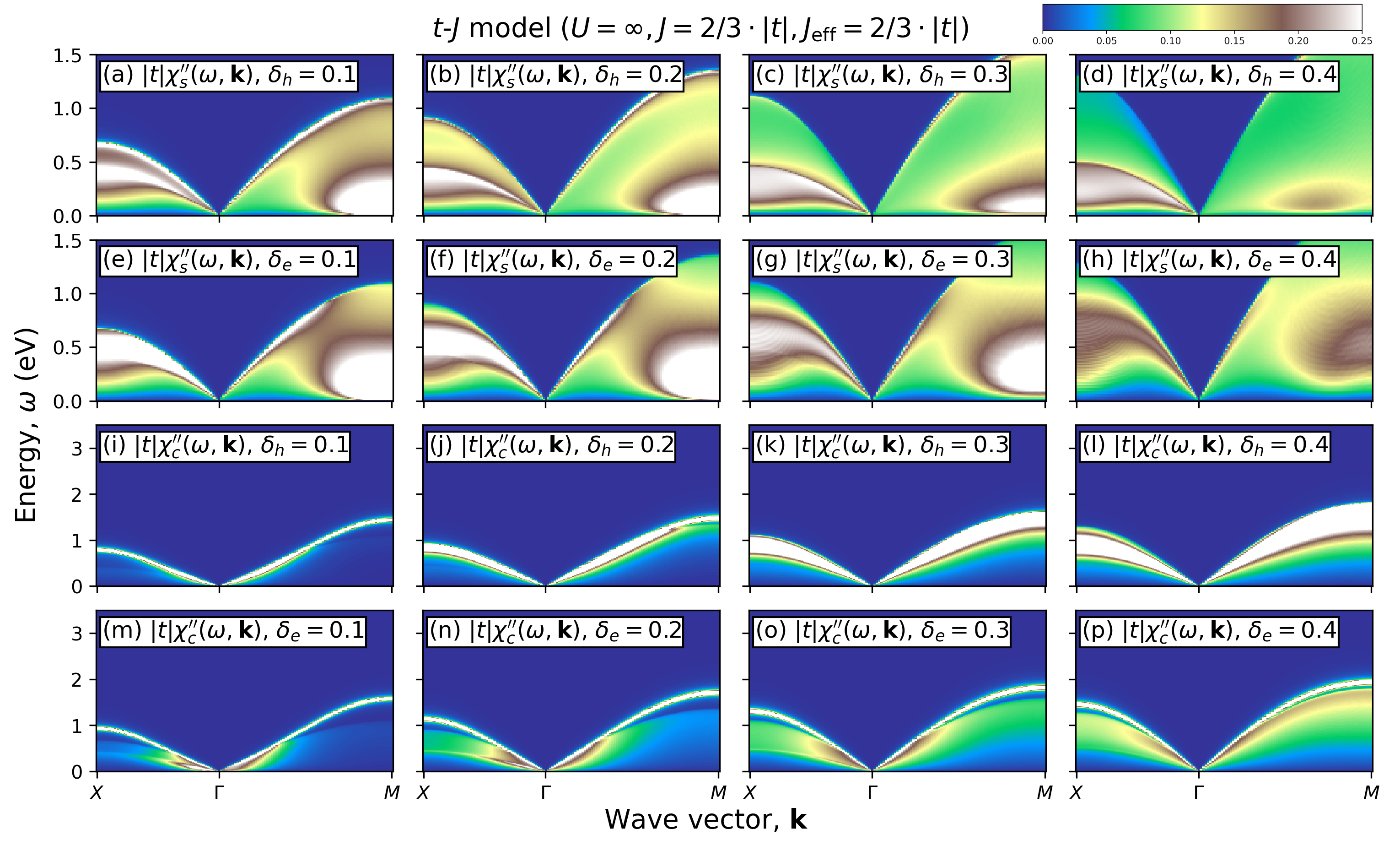}
  \caption{Summary of calculated VWF+$1/\mathcal{N}_f$ collective excitation spectra for the strong-coupling ($t$-$J$ model) limit ($t = -0.35\,\mathrm{eV}$, $t^\prime = 0.25 |t|$, $U = \infty$, $J = \frac{2}{3} |t|$). The plot is arranged in the same way as Fig.~\ref{fig:hubbard_panels}  and the color scale is consistent among panels.}
  \label{fig:tj_panels}
\end{figure*}

\subsection{$t$-$J$-$U$ model}
\label{sec:tju_model}

We now proceed to analysis of the $t$-$J$-$U$ model ($U = 12 |t|$ and $J = \frac{|t|}{3}$). In Fig.~\ref{fig:tju_panels} we display the calculated imaginary parts of dynamical spin- and charge susceptibilities for h- and e-doping. The results are organized in the same manner as those in Fig.~\ref{fig:hubbard_panels}. The dispersive magnetic peak appears on the h-doped side along the $\Gamma$-$X$ line, whereas $\Gamma$-$M$ excitations are highly incoherent [panels (a)-(d)]. Also, the magnetic mode rapidly dissolves in the continuum on the e-doped side for both analyzed high-symmetry Brillouin-zone directions [panels (e)-(h)]. All those features of the calculated spectra are qualitatively consistent with those obtained for the Hubbard model, which points towards essentially universal manifestation of the e-h asymmetry in magnetic dynamics.

The advantage of the $t$-$J$-$U$-model over either Hubbard- or $t$-$J$-model limits is that the scale of local correlations and kinetic exchange interactions are independently controlled by microscopic Hamiltonian parameters. Due to the same adopted value of $J_\mathrm{eff}$, the $t$-$J$-$U$-model dispersion of the coherent magnetic peak closely follows that of the Hubbard model [cf. Fig.~\ref{fig:hubbard_panels}(a)-(d)], even though the on-site interaction $U$ is larger by the factor of two. The physical consequence of increased on-site interaction is substantial narrowing of the continuum part of the spectrum and reduction of the charge-mode bandwidth, particularly close to half-filling.

\subsection{$t$-$J$ model as strong-coupling limit}
\label{sec:tj_model}

Finally, we address the strong-coupling limit by setting $U = \infty$ and $J = \frac{2}{3} |t|$.  Figure~\ref{fig:tj_panels} summarizes the calculated imaginary parts of the spin- and charge dynamical susceptibilities. A dispersive paramagnon-like peak with similar characteristics to those of Hubbard and $t$-$J$-$U$ models is visible only for the $\Gamma$-$X$ direction and h-doping [panels (a)-(d)]. Interestingly, the magnetic spectra on the e-doped side [panels (e)-(h)] are more robust and intense than those obtained for finite-$U$ models. 

The principal qualitative feature of the $t$-$J$ model, absent for weak- and intermediate coupling, is the emergence of sharp peaks at the magnetic continuum threshold, both along the $\Gamma$-$X$ and $\Gamma$-$M$ directions. Those are not unambiguously seen in spectroscopy, which suggests that the $t$-$J$ model might overestimate electronic correlations in the context of high-$T_c$ cuprates. The latter finding is also independently supported by former studies, showing that the $t$-$J$-$U$ model yields better overall agreement with experiment for those materials \cite{SpalekPhysRevB2017}.

The charge-excitation spectra, displayed in Fig.~\ref{fig:tj_panels}(i)-(p), are renormalized with respect to Hubbard and $t$-$J$-$U$ models, yet no qualitatively new behavior is observed.

\begin{figure*}
  \centering
    \includegraphics[width=1\linewidth]{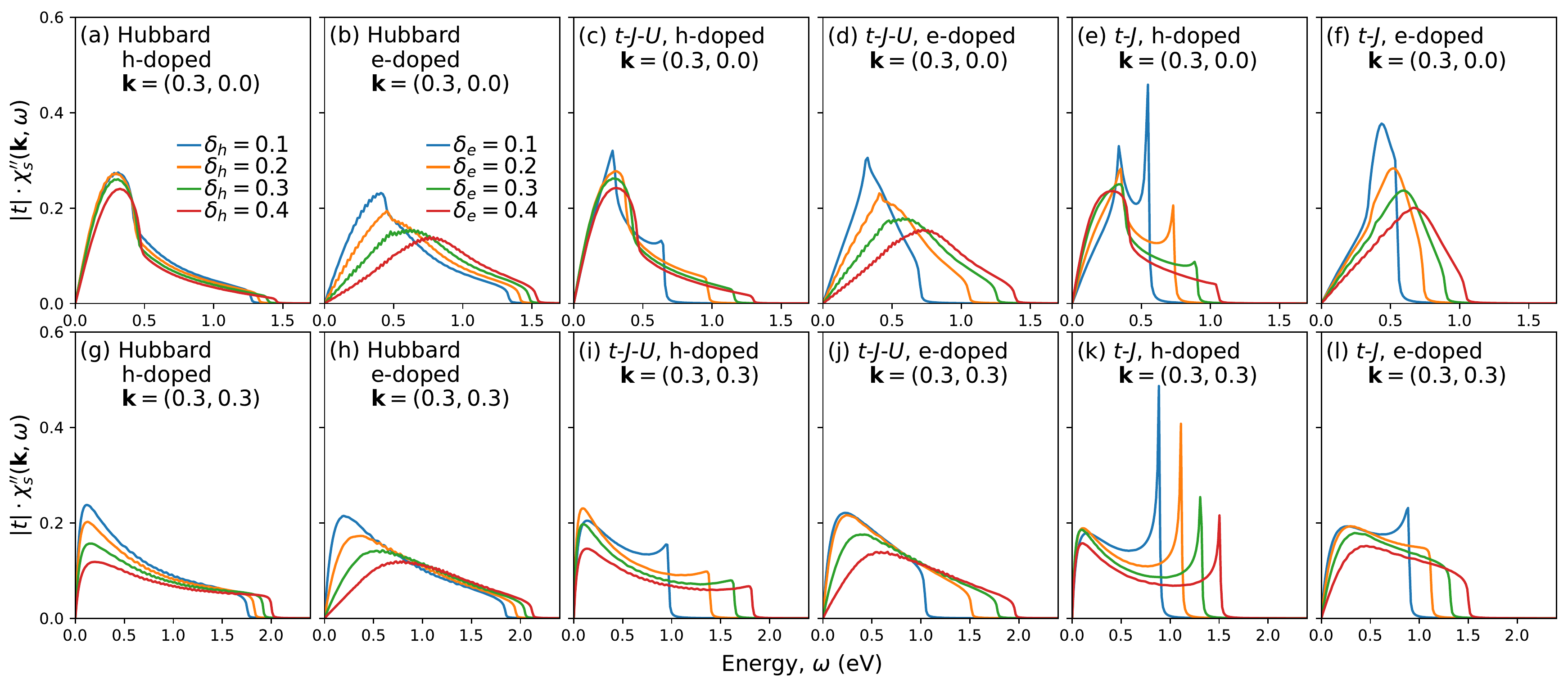}
  \caption{Energy profiles of the imaginary part of the dynamical spin susceptibility, $\chi^{\prime\prime}_s(\omega, \mathbf{k})$, obtained for the Hubbard, $t$-$J$-$U$, and $t$-$J$ models. The data are plotted for representative wave vectors, both on the h- and e-doping sides of the phase diagram. The parameters and models are indicated inside the panels (wave vectors are listed in the units $\frac{2\pi}{a}$ with $a$ being lattice spacing).}
  \label{fig:energy_scans}
\end{figure*}

\subsection{Structure of the paramagnon peaks}
\label{sec:spectral_intensities}

The relation between Hubbard, $t$-$J$-$U$, and $t$-$J$ models may be further characterized by comparing the energy profiles of the collective magnetic excitation spectra. This is carried out in Fig.~\ref{fig:energy_scans}, where we  display the imaginary part of dynamical spin susceptibility, plotted for all three models as a function of energy. Several representative doping levels and wave vectors are selected, with top and bottom panels corresponding to the anti-nodal and nodal directions, respectively (parameters are detailed inside the plot).

As noted above, paramagnon-like peaks emerge only for h-doping along the $\Gamma$-$X$ line [panels (a), (c), and (e)]. However, their lineshapes vary substantially depending on the on-site interaction strength, $U$. The magnetic peak contains a cusp for the $t$-$J$ model up to $\delta_h \lesssim 0.2$ [blue and orange curves in Fig.~\ref{fig:energy_scans}(e)] and $t$-$J$-$U$ model up to $\delta_h \lesssim 0.1$ [blue line in Fig.~\ref{fig:energy_scans}(c)]. This non-analytic feature is reminiscent of a Van Hove singularity in two-particle density of states, and thus \emph{is not} of resonant origin. Disentanglement of the resonant (paramagnon) part of the spectrum from non-resonant (continuum) excitations poses thus a major challenge in the strong-coupling regime. Those aspects will be addressed in detail elsewhere \cite{FidrysiakSpalek_unpublished}.

\section{Discussion}
\label{sec:summary}

\begin{figure*}
  \centering
    \includegraphics[width=1\linewidth]{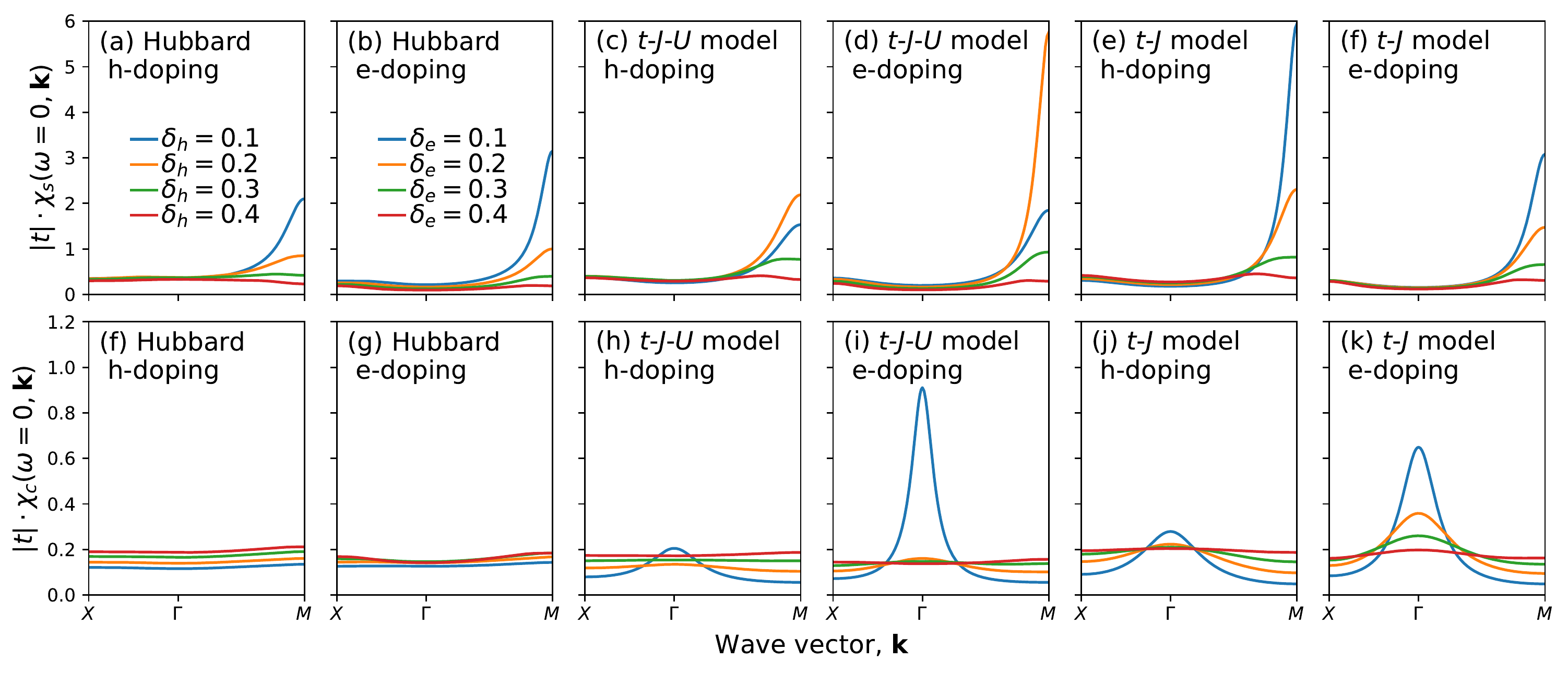}
  \caption{Calculated static spin- and charge susceptibilities for the Hubbard, $t$-$J$-$U$, and $t$-$J$ models, both on the h- and e-doped side of the phase diagram. The relevant parameters are detailed inside the panels. The obtained susceptibilities remain positive along the high-symmetry $X$-$\Gamma$-$M$ contour which signals stability of the high-temperature paramagnetic state against fluctuations.}
  \label{fig:stability}
\end{figure*}

In Sec.~\ref{sec:results} we presented the VWF+$1/\mathcal{N}_f$ characteristics of the collective spin- and charge excitations for the  Hubbard, $t$-$J$-$U$, and $t$-$J$ models of high-$T_c$ copper oxides, both on h- and e-doped side of the phase diagram. The microscopic parameters of those three Hamiltonians have been related to each other with the help of the effective exchange concept \cite{FidrysiakJMMM2021} so that Figs.~\ref{fig:hubbard_panels}-\ref{fig:tj_panels} can be directly compared. 

The main finding of the present analysis is that robust paramagnon peak emerges in the magnetic spectrum only along the anti-nodal $\Gamma$-$X$ direction for h-doping, whereas e-doped systems exhibit hardening of spin excitations and less coherent dynamics. On the other hand, coherent charge mode is found on both sides of the phase diagram. Those results are consistent among all models analyzed, and remain in agreement with RIXS experiments for multiple copper-oxide families. 

We also recall that the only term explicitly breaking the e-h symmetry in the Hamiltonian~\eqref{eq:tju-model} is the next-nearest-neighbor hopping $\propto t^\prime$ which controls single-particle excitations and is necessary match the high-$T_c$ copper oxide fermiology. Indirectly, $t^\prime$ impacts also the structure of continuum particle-hole excitations that contribute to the magnetic response. Thus, the obtained universal e-h asymmetry of the paramagnon-like spectrum for wide range of interaction parameters suggests that this particular effect is governed predominantly by the single-particle kinematics rather than by multi-particle correlations. 

We have also identified several principal differences between the spectra of the Hubbard, $t$-$J$-$U$, and $t$-$J$ models. As demonstrated in Sec.~\ref{sec:results}, continuum spin excitations undergo a substantial renormalization with increasing value of the on-site interaction, $U$. Moreover, in the strong coupling ($t$-$J$ model) limit, an additional sharp peak in the magnetic response appears at the continuum threshold. There are also more subtle differences in the spectral lineshapes. We observe that, on approaching the strong-coupling limit, the magnetic peaks evolve into the cusp-like structures characterized by discontinuous first derivative of magnetic intensity. This, in turn, points towards a non-negligible admixture of incoherent continuum excitations to magnetic signal.

We now comment on apparently opposite effect of the e-h asymmetry on the collective spin excitations and static magnetic order. In Appendix~\ref{appendix:stability} we have presented the VWF+$1/\mathcal{N}_f$ static susceptibilities. The magnetic response attains maximum at the antiferromagnetic $M$ point for both h- and e-doping, and is systematically larger on the e-side for the Hubbard and $t$-$J$-$U$ models. This signals an enhanced tendency towards AF ordering for e-doping, as seen in experiment \cite{ArmitageRevModPhys2010} (interestingly, this hierarchy is reversed in the strong coupling limit). Our present analysis of non-equilibrium quantities may be reconciled with the above result by noting that robust magnetic excitations emerge only along the $\Gamma$-$X$ direction, thus AF order and the paramagnons are governed by distinct regions of the Brillouin zone.

Finally, we note that electronic correlations beyond mean-field theory must be incorporated in order to accurately describe the magnetic dynamics, even for relatively small values of $U$. This has been demonstrated for the Hubbard-model case by comparison of the VWF+$1/\mathcal{N}_f$ and random-phase-approximation dynamical spin susceptibilities, and subsequently relating them both to experiment and quantum Monte-Carlo simulations \cite{FidrysiakPhysRevB2020,FidrysiakPhysRevB2021}. The present analysis suggests that both those correlation-related effects and non-resonant (incoherent) contributions are needed to consistently describe the magnetic spectra throughout the high-$T_c$ phase diagram.

\section*{Acknowldgments}

I thank Prof.~J.~Spa{\l}ek for discussion. This work was supported by Grants MINIATURA 5 No. DEC-2021/05/X/ST3/00666, OPUS No. UMO-2021/41/B/ST3/04070, and OPUS No. UMO-2018/29/ST3/02646 from Narodowe Centrum Nauki. Funding by ``Laboratories of the Young'' as part of the ``Excellence Initiative -- Research University'' program at the Jagiellonian University in Kraków is also acknowledged. 

\appendix

\section{Stability of the paramagnetic state}
\label{appendix:stability}

Here we analyze the stability of high-temperature paramagnetic state against both spin- and charge fluctuations for the models considered in the main text. In Fig.~\ref{fig:stability} we display calculated $X$-$\Gamma$-$M$ scans of spin [panels (a)-(f)] and charge [panels (g)-(k)] static susceptibilities, $\chi_s(\omega = 0, \mathbf{k})$ and $\chi_c(\omega = 0, \mathbf{k})$, respectively. The doping levels and models are indicated inside the panels, whereas the remaining parameters are the same as those taken in Sec.~\ref{sec:results}. Both spin and charge static susceptibilities remain positive in entire doping regime, showing that the paramagnetic ground state is stable against fluctuations on both h- and e-doped sides of the phase diagram.

There are several additional remarks that might be made, based on the analysis of Fig.~\ref{fig:stability}. First, for the Hubbard and $t$-$J$-$U$ models, the tendency towards antiferromagnetic ordering is stronger for e-doping, as evidenced by a larger value of static spin susceptibility close to the $M$ point, cf. panels (a)-(d). This is qualitatively consistent with experimental asymmetry of equilibrium magnetic phase diagram of the cuprates, with AF state being more robust on the e-doped side. Interestingly, the $t$-$J$ model yields opposite result [panels (e)-(f)], which favors finite-$U$ Hamiltonians for quantitative studies of high-$T_c$ systems. Second, for the $t$-$J$-$U$ model, the largest spin susceptibility magnitude is obtained for intermediate doping levels so that the ordering tendency is shifted away from half-filling.

On the other hand, charge susceptibility remains fairly featureless for all models and doping levels exceeding $0.1$. The enhancement of static charge response around the $\Gamma$ point should be regarded as an unphysical feature, since the Hamiltonian~\eqref{eq:tju-model} does not include long-range Coulomb interactions. The algebraic tail of the Coulomb potential is singular at $\mathbf{k} \rightarrow 0$, which is known to suppress the magnetic response in the vicinity of the Brillouin-zone center \cite{SpalekPhysRep2022}.

\section{Electron-hole symmetry of the variational space}
\label{appendix:correlator}

Here we show that the correlator~\eqref{eq:correlator} defines a variational space that is e-h symmetric. As a consequence, the obtained asymmetry of the collective excitations is an intrinsic feature of the $t$-$J$-$U$ model and is not induced by the particular choice of trial wave function.

We proceed by rewriting the correlator terms in second-quantization language:

\begin{align}
  |0\rangle_i{}_i\langle 0| &= (1 - \hat{n}_{i\uparrow}) (1 - \hat{n}_{i\downarrow}), \\
  |\sigma\rangle_i{}_i\langle \sigma| &= \hat{n}_{i\sigma} (1 - \hat{n}_{i\bar{\sigma}}) , \\
  |\sigma\rangle_i{}_i\langle \bar{\sigma}| &= \hat{a}^\dagger_{i\sigma} \hat{a}_{i\bar{\sigma}} , \\
  |d\rangle_i{}_i\langle d| &= \hat{n}_{i\uparrow} \hat{n}_{i\downarrow}, 
  \label{eq:correlator_second_quant}  
\end{align}

\noindent
where $\bar{\sigma} = \downarrow (\uparrow)$ for $\sigma = \uparrow (\downarrow)$. By applying the prescription~\eqref{eq:ph_transformation} and noting that $\hat{n}_{i\sigma} \rightarrow 1 - \hat{n}_{i\sigma}$, one arrives at

\begin{align}
  \label{eq:correlator_transformed}
  \hat{P}_i \rightarrow \lambda^0_{ i} |d \rangle_i {}_i\langle d | + \sum_{\sigma\sigma^\prime} \sigma\sigma^\prime \lambda^{\sigma\sigma^\prime}_{i} |\bar{\sigma}\rangle_i {}_i\langle \bar{\sigma}^\prime | + \lambda^d_{i} |0\rangle_i {}_i\langle 0 |.
\end{align}

\noindent
The transformed correlator~\eqref{eq:correlator_transformed} may be now brought to its original form~\eqref{eq:correlator} by rearranging the variational parameters as $\lambda^0_i \rightarrow \lambda_i^d$, $\lambda^{\sigma\sigma^\prime}_i \rightarrow \sigma \sigma^\prime \cdot \lambda_i^{\bar{\sigma}\bar{\sigma}^\prime}$, and $\lambda^d_i \rightarrow \lambda_i^0$, which concludes the reasoning. Note that the transformation rules for the single-particle part of the correlator, $\lambda^{\sigma\sigma^\prime}_i$, could be simplified by supplementing the usual e-h transformation with spin-flip operation, and including the spin-dependent phase factor in Eq.~\eqref{eq:ph_transformation}.

\end{document}